\documentclass[twoside]{mhd}

\usepackage{accents}
\usepackage{graphicx}

\mhdhead{40}{1}{1}	

\title{Growth rate of the initial magnetic energy in isotropic velocity field with non-Gaussian distribution}

\author{E.A. Illarionov\inst{1}\inst{,2}, D.D. Sokoloff\,\inst{1}\inst{,2}\inst{,3}}

\institute{Moscow State University, Moscow, Russia 
\and
Moscow Center of Fundamental and Applied Mathematics, Moscow, Russia
\and IZMIRAN, Moscow, Russia} 

\begin{document}
\maketitle


\begin{abstract}
We propose a method for including the effects of non-Gaussian velocity field distribution in the estimation of growth rate of the magnetic energy in a random flow with finite memory time. The method allows a reduction to the Gaussian case that was investigates earlier. For illustration we consider the multivariate Laplace distribution and compare it against the Gaussian one. 

\end{abstract}


\section*{Introduction}

Magnetic field growth in a flow of electrically conducting media (known as dynamo) is investigated from various viewpoint ranging from relative simple analytical models to direct numerical simulations and laboratory experiments (see e.g. \cite{Sokoloff2014} for a review). Of course,  main bulk of analytical results in the problem is obtained in the framework of most simple models, e.g. homogeneous, isotropic short-correlated Gaussian random flows. Any possibility to go beyond such oversimplifications looks attractive and instructive in a perspective of comparison with more realistic numerical simulations and laboratory dynamos, which are obviously nonhomogeneous, anisotropic, non-Gaussian and are remote from an instantaneous memory losses. Recent paper \cite{IS21} suggests a way how the finite memory and anisotropy  effects can be analytically investigated at least for the very first stage of dynamo action. Here we develop this method to include effects of non-Gaussian velocity field distribution. The main technical problem here is how to describe in an effective way a non-Gaussian velocity field distribution. We overcome the problem using
the idea suggested by \cite{Andrews1974}. To be specific, we consider here an isotropic case only, while the method is applicable to the anisotropic distribution as well.

\section{Governing equations}

We start from the induction equation for magnetic field $\bf H$ in a velocity field $\bf v$ that reads
\begin{equation}
    {{\partial {\bf H} \over {\partial t}} + ({\bf v} \nabla} {\bf H}) = ({\bf H } \nabla) {\bf v} - \eta \, {\rm curl} \, {\rm curl}\, {\bf H} \, .
    \label{ind}
\end{equation}
We assume that at the initial stage of magnetic field evolution 
the magnetic diffusivity $\eta$ can be neglected. Thus in the Lagrangian frame we obtain
from Eq.~(\ref{ind})
\begin{equation}
    {{d \bf H} \over {dt}} = A \bf H\, ,
\label{largange}
\end{equation}
where 
$A$ consists of
$ A_{ij} = \partial v_i/ \partial x_j$, taken at a given Lagrangian trajectory.

For a random velocity field  $\bf v$, the matrix $A$ should be considered as a random process. An effect of finite correlation time can be modelled as follows. Consider time instants $\tau$, $2\tau$, $3\tau$ and so on and assume that the process $A(t)$ is constant at intervals of length $\tau$ and its realizations at different intervals are statistically independent and identically distributed. This model is known as the innovation model (see e.g. \cite{Z1984}). Realization of the process $A(t)$ at the $i$-th time interval will be denoted as $\hat A_i$ using the hat symbol.

To fully characterize the process $A(t)$ it remains to set up the generating distribution for the random matrices $\hat A_i$. We proceed from the correlation tensor $R_{ij}({\textbf{x}-\textbf{y}}) = \langle v_i({\rm{\textbf{x}}})v_j({\rm{\textbf{y}}}) \rangle$ for statistically homogeneous and isotropic
velocity field. Following \cite{Batchelor1953} and \cite{Monin2007}  we consider
\begin{equation}
R_{ij}({\rm{\textbf{r}}}) = f(r)\delta_{ij} + \frac{r}{2}f'(r)\left ( 
\delta_{ij} - \frac{r_ir_j}{r^2} 
\right )\,,
\label{corr_comp}
\end{equation}
where $\textbf{r} = \textbf{x} - \textbf{y}$, $r=\|\textbf{r}\|$ and
$f$ is the longitudinal correlation function. We use  
$f(r) = (v^2/3)\exp(-3r^2/5l^2)$
that yields in
$\langle v_i({\rm{\textbf{x}}}) v_i({\rm{\textbf{y}}}) \rangle \sim v^2(1-(r/l)^2)$. Here
$l$ defines the scale of random vortexes, $v$ is
the the rms velocity of the flow \cite{SI15}. 

Given the correlation tensor $R_{ij}$ for the velocity field components one can obtain the correlation tensor for derivatives of the velocity field. It can be obtained by taking second order 
partial derivatives
$\partial^2/\partial x_m\partial y_n$ of
$R_{ij}(\rm{\textbf{r}})$ and computing limit $\rm{\textbf{r}}\to 0$. The resulting tensor of size $9\times9$ will be denoted as $B$.

For simplicity we assume $\langle {\bf v} \rangle=0$ and thus mean of the matrices $\hat A_i$ vanishes. The details of selecting an appropriate probability distribution that results in zero mean and covariance matrix $B$ are the key part of this work and will be considered in the next section. Whatever the distribution is fixed, the mean magnetic field is obtained as follows:
\begin{equation}
    \langle{\bf H}\rangle(n\tau) = \langle \exp ( {\hat A}\tau) \rangle^n {\bf H}_0\, ,
\label{vector_avr} 
\end{equation}
where $\hat A$ stands for any of the identically distributed matrices $\hat A_i$. As a result, growth rate for $\langle{\bf H}\rangle$ is defined by the leading eivenvalue of the mean matrix exponential $\langle \exp ( {\hat A}\tau) \rangle$.

An equation for the second statistical moment $\|{\bf H}\|^2$ requires an extension of the Eq.~\ref{largange}. Let $Z_{ij} = H_i H_j$ and consider the vector ${\bf Z} = (Z_{11}, Z_{12}, ..., Z_{33})^T$ of length $9$. Computing  derivatives $\partial Z_{ij} /\partial t = \partial (H_i H_j) /\partial t$ and taking into account Eq.~\ref{largange} we arrive to the extended equation
\begin{equation}
    \frac{d \bf Z}{d t} = {\cal A}(t) {\bf Z}\, ,
\label{vector_ext}    
\end{equation}
where the matrix ${{\mathcal A}}(t)$ is related to the matrix $A(t)$ as follows:
\begin{equation}
   {\cal  A}_{k(i-1)+j, k(l-1)+m} = {A}_{il}\delta_{jm} + {A}_{jm}\delta_{il} \, .
   \label{A_hat}
\end{equation}
Similar to Eq.~\ref{vector_avr}, growth rate of the mean vector $\langle \bf Z \rangle$ is defined by the leading
eigenvalue of the mean matrix exponential
$\langle\exp({\hat{\mathcal A}} \tau)\rangle$. Since $Z_{ii}  = H_i^2$, the same eigenvalue also defines the growth rate of $\langle \|{\bf H}\|^2 \rangle$.

To estimate the leading eigenvalue we approximate the mean matrix exponential with a Taylor series and consider the characteristic equation
\begin{equation}
\textrm{det}\left(I(1-\lambda) + \frac{1}{2}\langle {\hat{\mathcal A}}^2 \rangle\tau^2 +  \frac{1}{4!}\langle {\hat{\mathcal A}}^4 \rangle\tau^4 + ...\right) = 0 \, .
\label{char_eq}
\end{equation}
Eq.~\ref{char_eq} allows approximation of eigenvalues with any desired accuracy both analytically and numerically, however, computation of higher statistical moment of the random matrix might be complicated. In the next section we demonstrate, that many practically interesting distributions may be reduced to the Gaussian one, for which the result has been already obtained in \cite{IS21}.
In terms of the Strouhal number $s=\tau v/l$, an $O(s^8)$ approximation of the leading eigenvalues reads 
\begin{equation}
    \lambda(s) = 1 + 2s^2 + \frac{9}{10}s^4 + \frac{149}{500}s^6 + O(s^8) \, .
\label{roots}
\end{equation}
Proceeding from the definition of the growth rate of the $p$-th statistical moment
\begin{equation}
    \gamma_p = 
    \lim\limits_{n\to\infty}\frac{1}{2pn\tau}\ln\langle  \|{\bf H}(n\tau)\|^p \rangle \, ,
\end{equation}
growth rate of the second moment ($p=2$) can be evaluated as $\gamma_2 = (1/4\tau)\ln \lambda$.

\section{Main results}

The Gaussian distribution for matrices $\hat A$ is a convenient however not necessary assumption for the given mean value and covariance matrix. For example, one could consider the multivariate Laplace or Student distribution. Our main result is to demonstrate that for a broad family of distributions, the leading eigenvalue for the mean matrix exponential is obtained from Eq.~\ref{roots}, which is computed for the Gaussian case. This means that the lengthy algebra that was behind the Eq.~\ref{roots} can be simplified substantially for new distributions.

Consider 
$\hat A = \sqrt{Y}X$, 
where X has a multivariate Gaussian distribution and Y is an independent non-negative random value. This model is known as a multivariate scaled mixture of Gaussians (see e.g. \cite{Andrews1974}). The model has several important special cases. In particular, it can be shown (\cite{Eltoft2006}) that if $Y$ is an exponential variable with probability density function (pdf) $p_Y(x) = (1/\beta)\exp(-x/\beta)$ for $\beta > 0$,
while $X$ has Gaussian distribution with zero mean (for simplicity) and covariance matrix $\Sigma$, then $\sqrt{Y}X$ has the multivariate Laplace distribution with pdf
\begin{equation}
    p({\bf x}) = \frac{1}{\sqrt{(2\pi)^{d}|\Sigma|}}\frac{2}{\beta}\frac{K_{(d/2-1)}\left(\sqrt{\frac{2}{\beta}{\bf x'}\Sigma^{-1}{\bf x}}\right)}{\left( \sqrt{\frac{\beta}{2}{\bf x'}\Sigma^{-1}{\bf x}}\right)^{d/2-1}} \, .
\end{equation}
$d=3$ is space dimension and 
$K_{1/2}(x) = \sqrt{\pi/2x}e^{-x}$
is the modified Bessel function of the second kind, ${\bf x'}$ is the transposition of {\bf x}. 

Now we turn to the Eq.~\ref{char_eq} taking into account that random matrices are from the multivariate scaled mixture of Gaussians. The characteristic equation becomes 
\begin{equation}
\textrm{det}\left(I(1-\lambda) + \frac{1}{2}\langle Y\rangle\langle {\hat{\mathcal X}}^2 \rangle s^2 +  \frac{1}{4!}\langle Y^2\rangle\langle {\hat{\mathcal X}}^4 \rangle s^4 + ...\right) = 0 \, .
\label{char_eq2}
\end{equation}
Here ${\hat{\mathcal X}}$ is computed from $X$ according to Eq.~\ref{A_hat}. 
It follows from \cite{IS21} that in isotropic case the leading eigenvalue can be found as $\langle \lambda(\sqrt{Y}s)\rangle$,
where $\lambda(\cdot)$ is defined in Eq.~\ref{roots}. We arrive to the general equation, which is valid in isotropic case:
\begin{equation}
    \lambda(s) = 1 + 2\langle Y\rangle s^2 + \frac{9}{10}\langle Y^2\rangle s^4 + \frac{149}{500}\langle Y^3\rangle s^6 + O(s^8) \, .
\label{roots}
\end{equation}
Note that in anisotropic velocity field eigenvalues should be explicitly derived from Eq.~\ref{char_eq2} and we do not consider it in this work.

\begin{figure}
\centering
\includegraphics[width=0.75\textwidth]{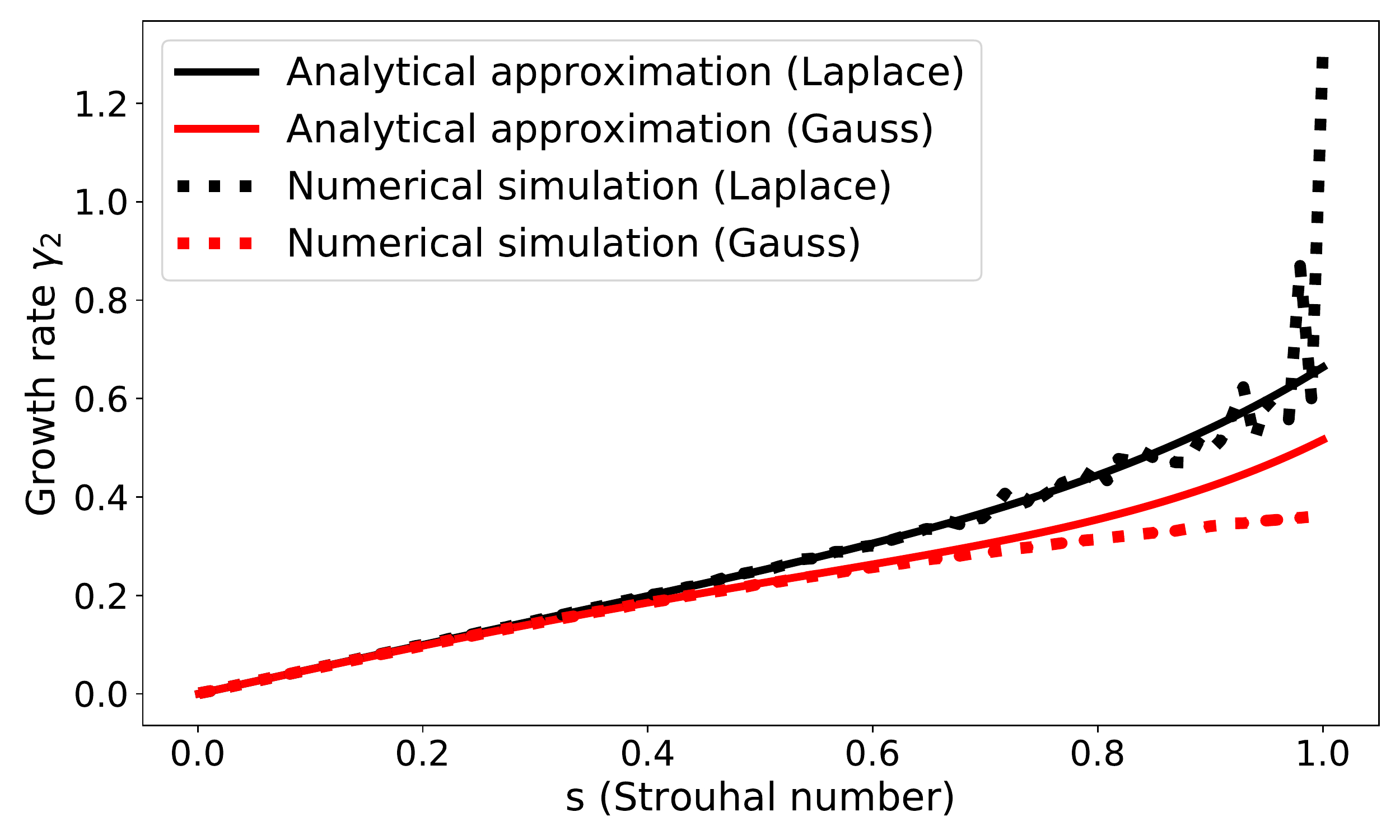}
\caption{Growth rate of the second moment of magnetic field in case of Gaussian and Laplace distribution of velocity field derivatives and for various Strouhal numbers. Solid lines correspond to analytical approximation of the growth rates, dotted lines show the results of numerical simulation.}
\label{fig:laplace}
\end{figure}

As a particular case consider $Y$ distributed exponentially with $\beta=1$ (it corresponds to the Laplace distribution for $\hat A$). Then $\langle Y^n\rangle = n!$ and we obtain 
\begin{equation}
    \lambda(s) = 1 + 2 s^2 + \frac{9}{5}s^4 + \frac{447}{250} s^6 + O(s^8) \, .
\label{roots}
\end{equation}
In terms of the growth rate it corresponds to (omitting the scale factor $v/l$)
\begin{equation}
    \gamma_2 = \frac{1}{2}s -\frac{1}{20}s^3 +  \frac{641}{3000}s^5 +  O(s^7) \, .
\label{roots}
\end{equation}
In order to compare the above analytical approximation with a numerical experiment, we simulate $10^5$ realizations of random matrices $\hat A$ following the multivariate Laplace distribution and repeat it independently for each Strouhal number $s$ from the interval (0, 1). Then for each $s$ we compute sample averaged matrix exponential $\langle\exp({\hat{\mathcal A}}s)\rangle$,
find numerically its leading eigenvalue $\lambda$ and finally obtain $\gamma_2=(1/4s)\ln\lambda$. In Fig.~\ref{fig:laplace} we demonstrate that the numerical simulation becomes unstable for $s>0.6$ and matches with the analytical approximation for $s<0.6$. Similar results were obtained for the case of Gaussian distribution of random matrices. Also in Fig.~\ref{fig:laplace} we compare the cases of Laplace and Gaussian distributions and as expected find that the Laplace distribution results higher growth rates due to heavier tails of the distribution.

\section{Conclusions}

We suggested a generalization of the method to obtain magnetic energy growth rates in a finite memory random flows to a non-Gaussian velocity distribution. We remind here that a random flow with instantaneous memory losses has to be Gaussian what is not the case for the finite memory flows. It is why the problem under consideration looks natural and instructive. Fortunately, we obtained that the effects of non-Gaussianity are moderate and the simplest results for short-correlated Gaussian flows remain instructive in non-Gaussian perspective. In this sense the results can be considered as a generalisation of that one obtained in \cite{IS21}.

\section*{Acknowledgements}
Authors acknowledge the financial support of the Ministry of Education and Science of the Russian Federation as part of the program of the Moscow Center for Fundamental and Applied Mathematics under the agreement № 075-15-2022-284. DS acknowledges the financial support from the Basis Foundation under grant № 21-1-1-4-1.

\bibliographystyle{mhd}
\bibliography{literature}
\clearpage




\lastpageno
\end{document}